\documentclass[journal,letterpaper,twocolumn,final,10pt]{IEEEtran}

\usepackage{amssymb}
\usepackage{amsthm}
\usepackage{amsmath}
\usepackage{amsfonts}
\usepackage{dsfont}
\usepackage{mathrsfs}
\usepackage{pifont}
\usepackage{amssymb}
\usepackage{verbatim}
\usepackage{upgreek}
\usepackage[final]{graphicx}
\usepackage{epstopdf}
\usepackage{algorithmic}
\usepackage{algorithm}
\usepackage{epsfig}
\usepackage{eucal}
\usepackage{cite}
\usepackage{multirow}
\usepackage{hyperref}
\usepackage{color}
\ifCLASSINFOpdf
\else
   \usepackage[dvips]{graphicx}
\fi
\usepackage{url}

\usepackage{subfigure}

\newcommand{\bbm}{\begin{bmatrix}}
\newcommand{\ebm}{\end{bmatrix}}

\newcommand{\bit}{\begin{itemize}}
\newcommand{\eit}{\end{itemize}}

\newcommand{\ben}{\begin{enumerate}}
\newcommand{\een}{\end{enumerate}}

\newcommand{\bdesc}{\begin{description}}
\newcommand{\edesc}{\end{description}}

\newcommand{\bea}{\begin{array}}
\newcommand{\eea}{\end{array}}

\newcommand{\tr}{\mbox{\rm Tr}\, }

\newcommand{\beqa}{\begin{eqnarray}}
\newcommand{\eeqa}{\end{eqnarray}}

\newcommand{\Comment}[1]{}

\def\R{{\mathds R}}

\def\cN{\mbox{$\CMcal N$}}

\def\cO{\mbox{$\mathcal O$}}

\newcommand{\be}{\begin{equation}}

\newcommand{\ee}{\end{equation}}

\newcommand{\boe}{{\mbox{\boldmath $e$}}}

\newcommand{\bm}{{\mbox{\boldmath $m$}}}

\newcommand{\bz}{{\mbox{\boldmath $z$}}}

\newcommand{\bA}{{\mbox{\boldmath $A$}}}

\newcommand{\bC}{{\mbox{\boldmath $C$}}}

\newcommand{\bM}{{\mbox{\boldmath $M$}}}

\newcommand{\bZ}{{\mbox{\boldmath $Z$}}}

\newcommand{\bSigma}{{\mbox{\boldmath $\Sigma$}}}

\newcommand{\dmax}{\begin{displaystyle}\max\end{displaystyle}}

\newcommand{\test}{\mbox{$
\begin{array}{c}
\stackrel{ \stackrel{\textstyle H_1}{\textstyle >} }{
\stackrel{\textstyle <}{\textstyle H_0} }
\end{array}
$}}

\begin{document}

\title{Design and Experimental Assessment of Detection Schemes for Air Interface Attacks \\ in Adverse Scenarios}

\author{D. Orlando, I. Palamà, S. Bartoletti, G. Bianchi, and N. Blefari Melazzi
\thanks{This work was supported by the European Union's Horizon 2020 research and innovation programme under Grant no. 871249.}
\thanks{D. Orlando is with the Universit\`a degli Studi ``Niccol\`o Cusano,'' 00166 Roma, Italy (e-mail: danilo.orlando@unicusano.it).}
\thanks{G. Bianchi, N. Blefari Melazzi, and I. Palamà are  with  Department of Electronic Engineering, 
Univ. of Roma Tor Vergata and CNIT, Via del Politecnico, 1 00133, Rome, Italy (email: 
\{giuseppe.bianchi, blefari, ivan.palama\}@uniroma2.it).}
\thanks{S. Bartoletti is with the National Research Council of Italy, CNR-IEIIT and CNIT, viale Risorgimento 2, 
40136 Bologna, Italy (e-mail: stefania.bartoletti@cnr.it).}}

\markboth{Journal of \LaTeX\ Class Files, Vol. XX, No. X, September 2020}
{Shell \MakeLowercase{\textit{et al.}}: Bare Demo of IEEEtran.cls for IEEE Journals}
\maketitle

\begin{abstract}
In this letter, we propose three schemes designed to detect attacks over the air interface in cellular 
networks. These decision rules rely on the generalized likelihood ratio test, and are fed by data that can be 
acquired using common off-the-shelf receivers. In addition to more classical (barrage/smart) noise jamming attacks, we 
further assess the capability of the proposed schemes to detect the stealthy activation of a rogue base station. 
The evaluation is carried out through an experimentation of a LTE system concretely reproduced using 
Software-Defined Radios. Illustrative examples confirm that the proposed schemes can effectively detect air 
interface threats with high probability.
\end{abstract}


\IEEEpeerreviewmaketitle

\section{Introduction}
Thirty years ago, wireless cellular networks were threatened only by a handful of resource-rich opponents equipped with 
tailored instrumentation. Today, any tech-savvy can acquire an ultra-cheap software-defined radio board, instruments it 
with an open-source software implementation of the LTE/5G protocol stack, and threaten the network with a plethora of offensive actions. 
Most of the radio interface attacks are grounded on the suitable combination of targeted jamming signals 
\cite{SmartJammingAttacksRev,Guide5Gsec} 
to force the User Equipment (UE) to abandon the (interfered) legitimate operator signal and make it connect 
to a (fake) Rogue Base Station (RBS) controlled by the adversary. 
At this time, the opponent may suitably spoof unauthenticated protocol/signaling messages so as to steer the 
victim into: Man-In-The-Middle scenarios such as downgrade/bidding-down attacks \cite{shaik1, shaik2};
location privacy threats such as tracking or International Mobile Subscriber Identity (IMSI) catching \cite{p1,PALAMA2021108137};
device capability information gathering \cite{mnmap}, and so on \cite{8310016}.

The evolving 3GPP standards have devised solutions to most of these threats. The latest example is the public-key-based 
SUbscriber Concealed Identity (SUCI) recently standardized \cite{suci} in 5G to ultimately solve the IMSI catching problem. 
However, attackers can leverage practical concerns that force operators to slowly and incrementally deploy 
next-generation cellular technologies (in fact, as we write, 2G systems are still active!) 
and convince the UE to believe that the only base station available in a coverage area is a fake one 
implementing a past generation standard, thereby circumventing the new protections.

This paper takes the lead from the remark that air interface attacks to cellular networks can be 
thwarted only by developing compelling techniques and systems which detect the early-warning signs of their appearance, 
namely (possibly smart) jamming and rogue base station activities, etc. Our specific contribution is twofold.
First, we devise three decision schemes based upon the Generalized Likelihood Ratio Test (GLRT)
to approach the above detection problems.
While the baseline theory is well-known \cite{KayBook,muirhead2009aspects}, our novelty consists in 
the tailored design and adaptation to the specific detection problem at hand, namely, jamming and RBS signals detection. 
The idea behind our approach consists of adaptively monitoring, within a preassigned temporal sliding window, a number 
of physically observable quantities that can be gathered from commodity receivers. 
The three designs do instead differ in the subset of parameters whose variations are representative of the effects of
malicious actions over the air interface.
%%%The three designs do instead differ in the set of parameters characterizing the assumed data distribution that are subject 
%%%to possible variations caused by the malicious agent actions.
Second, our contribution also focuses on the experimental assessment of the proposed architectures 
in a real-world experimental playground setup based upon LTE by using Software-Defined Radios 
respectively instrumented as a jamming device or 4G RBS, as a 4G UE, and as 4G eNB. In the experiments, 
we test our change detectors using the available observables such as the Signal-to-Noise Ratio (SNR) and/or 
noise power level measurements that can be retrieved by means of common mobile applications.
The corresponding decision rules (including well-known change detectors) are compared in 
terms of detection probability ($P_\textrm{d}$) over real recorded data. In addition to insights on the collected 
measurements, these experiments allow us to confirm that the temporally white Gaussian assumption 
on the measurement noise, which is suitable for analytical tractability of the problem, has a limited impact 
on the detection performance in real world where, clearly, the measurement noise is not deemed to 
follow such a tractable model.

%The remainder of this letter is organized as follows: Section II contains the problem formulation and the proposed 
%threat detection architectures, whereas Section IV is devoted to the numerical analysis and discussion. 
%Finally, Section V concludes this letter outlining future research tracks. The analytical derivations are confined to the appendices.

%%%%\subsection{Notation}
%%%%In what follows, vectors and matrices are denoted by boldface lower-case and upper-case letters, respectively.
%%%%Symbols $(\cdot)^T$, $\tr(\cdot)$, and $\det(\cdot)$ denote the the transpose, the trace, and the determinant, 
%%%%respectively, of the matrix argument.
%%%%As to the numerical sets, $\R^{N\times M}$ 
%%%%is the Euclidean space of $(N\times M)$-dimensional real matrices (or vectors if $M=1$).
%%%%$\bOne$ denotes the vector with unit entries whose size depends on the the context.
%%%%Finally, we write $\boldsymbol{x}\sim\cN_N(\boldsymbol{m},\boldsymbol{M})$ if $\boldsymbol{x}$ is an 
%%%%$N$-dimensional Gaussian vector with mean $\boldsymbol{m}$ and positive definite covariance matrix $\boldsymbol{M}$.

\section{Problem Statement and Design Issues}
Let us consider an operating scenario where the UE is connected to a genuine base station and a malicious agent is interested
in acquiring sensitive data through a RBS. To this end, the latter exploits three different strategies based on either: 
a Barrage Noise-like Jammer (BNLJ) \cite{poisel2011modern}, 
a (narrowband) Smart Noise-like Jammer (SNLJ) \cite{poisel2011modern}, 
or the interfering signals generated by the RBS itself.
A BNLJ places noise energy across the entire width of
the frequency spectrum used by the target communication systems 
(or, at least, simultaneously into a broad range of frequencies), whereas the considered
SNLJ attacks noncontiguous narrow portions of the target spectrum at a given time (this kind of attack belongs to the class
of the so-called partial-band noise jammers) \cite{poisel2011modern}.
The effects of these actions
consist in an increase of the interference level within the UE receiver and a consequent reduction of the SNR.
In this scenario, we assume that the UE is capable of collecting and processing the related measurements of 
SNR, instantaneous noise power level,
and average noise power level.\footnote{Actually, there exists a plethora of applications that enable the collection of 
these measurements such as Netmonster \copyright, Network Signal Guru \copyright, or Network Cell Info \copyright.}
Then, at each time instant, such measurements are organized to form $N$-dimensional (with $N=1,2,3$)
vectors\footnote{
Vectors and matrices are denoted by boldface lower-case and upper-case letters, respectively.
Symbols $(\cdot)^T$, $\tr(\cdot)$, and $\det(\cdot)$ denote the the transpose, the trace, and the determinant, 
respectively, of the matrix argument.
As to the numerical sets, $\R^{N\times M}$ 
is the Euclidean space of $(N\times M)$-dimensional real matrices (or vectors if $M=1$).
$\boe$ denotes the vector with unit entries whose size depends on the the context.
Finally, we write $\boldsymbol{x}\sim\cN_N(\boldsymbol{m},\boldsymbol{M})$ if $\boldsymbol{x}$ is an 
$N$-dimensional Gaussian vector with mean $\boldsymbol{m}$ and positive definite covariance matrix $\boldsymbol{M}$.
}, $\bz_k$ say, that, due to the measurement noise, are random and assumed to follow the multivariate Gaussian 
distribution with a given mean and unknown positive definite covariance matrix, $\bM$ say. The UE observes a finite temporal 
series of consecutive measurements to understand whether or not is under attack.
Therefore, let us focus on a temporal window of duration $K$ and denote by 
$\bZ=[\bz_1,\ldots,\bz_K]\in\R^{N\times K}$ the overall data matrix associated with this 
window.
The effects of the malicious agent translate into a variation of the data distribution parameters within the observation window.
To be more precise, there exists a time instant represented by an unknown index $K_1\in\{1,\ldots,K\}$ 
after which data might have a different mean and/or covariance matrix. Therefore, detecting this kind of malicious actions naturally
amounts to a change detection problem.

In what follows, we consider three different problems depending on the different situations that may occur.
The first problem assumes that only the covariance matrix changes after the instant $K_1$ while the means are generally different.
Otherwise stated, the effect of the jammer consists in an increase of the measurement variances only.
Thus, from a more formal standpoint, it can be formulated in terms of the following binary hypothesis test, namely
\be 
\left\{
\begin{aligned}
& H_0:
\begin{cases}
\bz_{1},\ldots, \bz_{K_1} \sim \cN_N(\bm_1,\bSigma),
\\
\bz_{K_1+1},\ldots, \bz_{K} \sim \cN_N(\bm_2,\bSigma),
\end{cases}
\\
& H_1:
\begin{cases}
\bz_{1},\ldots, \bz_{K_1} \sim \cN_N(\bm_1,\bSigma_1),
\\
\bz_{K_1+1},\ldots, \bz_{K} \sim \cN_N(\bm_2,\bSigma_2).
\end{cases}
\end{aligned}
\right.
\label{eqn:generalProblem}
\ee
Notice that the mean variation under $H_0$ accounts for changes triggered by possible environment (unintentional) interference.
Closed-form expression of the GLRT for the above problem cannot be obtained (at least to the best of authors' knowledge) due
to the maximization over $K_1$ which can be conducted using a suitable search grid. Following the 
lead of \cite{muirhead2009aspects}, it is possible to show that the logarithm of the GLRT (Log-GLRT) can be 
recast as (for the reader ease in Appendix \ref{App:derivationNCD} we include a sketch of the proof)
\begin{multline}
\dmax_{K_1} \left\{-\frac{K_1}{2}\log\det( \bA_1 / K_1 ) -\frac{K_2}{2}\log\det( \bA_2 / K_2 )\right\}
\\
+\dmax_{K_1}\left\{\frac{K}{2}\log\det( (\bA_1+\bA_2) / K )\right\}\test \eta,
\label{eqn:naiveCD}
\end{multline}
where $\eta$ is the threshold\footnote{Hereafter, we denote by $\eta$ the generic detection threshold.} 
to be set according to the probability of false alarm ($P_\textrm{fa}$), 
$K_2=K-K_1$, and\footnote{The value of $K$ is selected
to ensure a reasonable compromise between computational load and minimum number of samples required 
to intercept the entire change in data (which depends on the finite bandwidth of the signal generators).}
$\bA_i=\bZ_i\bZ_i^T - K_i \widehat{\bm}_i\widehat{\bm}_i^T$, $i=1,2$, with 
$\bZ_i=[\bz_{(i-1)K_1+1},\ldots,\bz_{(i-1)K_1+K_i}]$ and $\widehat{\bm}_i=(1/K_i)\bZ_i\boe$.
In the following, this detector will be referred to as Naive Change Detector (NCD).

The second problem is a modification of \eqref{eqn:generalProblem} assuming that under $H_0$ 
both the mean and the covariance matrix of the $\bz_k$s are constant with $k$,
namely, we neglect possible changes due to other
interference sources. Thus, we can write
\be 
\left\{
\begin{aligned}
& H_0: \bz_{1},\ldots, \bz_{K} \sim \cN_N(\bm,\bSigma),
\\
& H_1:
\begin{cases}
\bz_{1},\ldots, \bz_{K_1} \sim \cN_N(\bm_1,\bSigma_1),
\\
\bz_{K_1+1},\ldots, \bz_{K} \sim \cN_N(\bm_2,\bSigma_2).
\end{cases}
\end{aligned}
\right.
\label{eqn:sameMeanVarianceH0}
\ee
As shown in Appendix \ref{App:derivationMNCD}, the Log-GLRT for this new problem is given by
\begin{multline}
\dmax_{K_1} \Bigg\{-\frac{K_1}{2}\log\det( \bA_1 / K_1 ) 
-\frac{K_2}{2}\log\det( \bA_2 / K_2 )
\\
+ \frac{K}{2}\log\det( \bA_0 / K )\Bigg\}\test \eta,
\label{eqn:modifiedNCD}
\end{multline}
where $\bA_0=\bZ\bZ^T-K\widehat{\bm}\widehat{\bm}^T$ with $\widehat{\bm}=(1/K)\bZ\boe$, and
we will refer to it as Modified NCD (MNCD).

The last problem that is worth to be investigated moves the effects of the attacks on the mean only, 
namely,\footnote{Notice that problem \eqref{eqn:generalProblem}
encompasses \eqref{eqn:sameMeanVarianceH0} and \eqref{eqn:sameVarianceOnly}
as special cases. As a consequence, detector \eqref{eqn:naiveCD} is more general than \eqref{eqn:modifiedNCD}
and \eqref{eqn:SdP} since the latter, at design stage, exclude any mean variation (under $H_0$) induced by 
environment interference.}
\be 
\left\{
\begin{aligned}
& H_0: \bz_{1},\ldots, \bz_{K} \sim \cN_N(\bm,\bSigma),
\\
& H_1:
\begin{cases}
\bz_{1},\ldots, \bz_{K_1} \sim \cN_N(\bm_1,\bSigma),
\\
\bz_{K_1+1},\ldots, \bz_{K} \sim \cN_N(\bm_2,\bSigma).
\end{cases}
\end{aligned}
\right.
\label{eqn:sameVarianceOnly}
\ee
Thus, the final Log-GLRT considered here has the following expression
\be
\dmax_{K_1} \left\{-\log\det( \bA_1 + \bA_2 ) + \log\det( \bA_0 )\right\}\test \eta
\label{eqn:SdP}
\ee
and it has been derived in \cite{AnomalyDet}, where, with application to location security, it is called Spoofer Detector (SpD).

In the next section, we assess the performance of \eqref{eqn:naiveCD}, \eqref{eqn:modifiedNCD}, and \eqref{eqn:SdP} using real recorded
data under the attack of either a BNLJ, a SNLJ, or a RBS.

\section{Performance Assessment and Comparisons}
\label{Sec:Performance}
We now describe the experimental setup and the three operating scenarios for the detection performance assessment.
\vspace{-0.7cm}
\subsection{Experimental Setup}
\begin{figure}[h!]
\centering
\includegraphics[width=5cm]{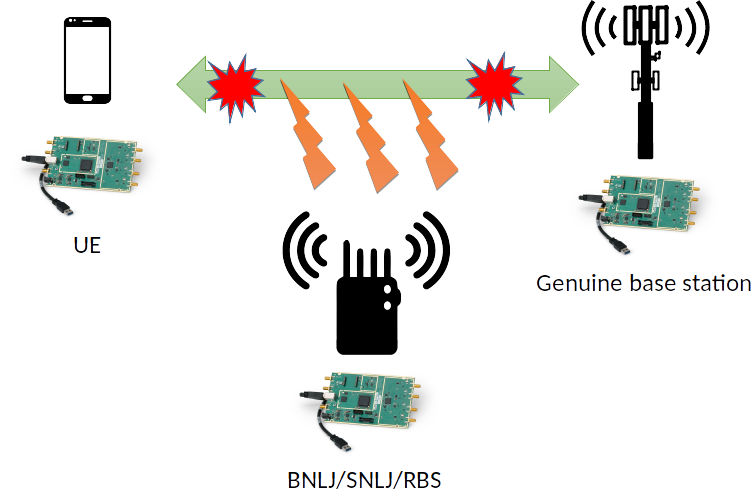}
\caption{Experimental setup.}
\label{fig:operatingScenario}
\end{figure}

The herein considered experimental setup encompasses three Ettus USRP B210 devices that can be tuned over 
a wide radio frequency range, from 70 MHz to 6 GHz, and, hence, can cover all the LTE frequency bands. 
The three radio transceivers are used to emulate a legitimate base station (eNB), the UE, and the 
attacker (see Figure \ref{fig:operatingScenario}).
These entities are stationary and placed at a distance of about $1$ meter from each other 
(due to legal power limits) within the laboratory
(indoor environment) where other electromagnetic sources are present and operative.
The communications between the UE and eNB use a bandwidth of 5 MHz (25 resource blocks) on the LTE band no. 7 ($2600$ MHz), 
which, due to its electromagnetic nature, minimizes the interference with the public 4G network. The parameter setting of the eNB, 
i.e., Mobile Country Code, Mobile Network Code, and Tracking Area Code (TAC), is such as to avoid 
any conflict with the public land mobile networks in Italy, the Physical Cell Identity is set to 0 while TAC is equal to 1.
The jammer disturbs the downlink communication emulating an environment in which the UE cannot receive signals 
from the eNB. Finally, the transmitted powers are such that the SNR at the receiver is about $20$ dB
while the Jamming-to-Noise Ratio (JNR) is equal to either $20$ or $25$ dB leading to a Jamming-to-Signal Ratio (J/S) of
$0$ and $5$ dB, respectively.
On the software side, we exploit the OpenAirInterface (OAI) project that provides a full real-time 
indoor/outdoor experimental 3GPP-compliant LTE implementation \cite{OpenAirInterface1}. 
More importantly, it is a reference platform in 
the context of 5G research and includes: OAI Radio Access Network (OAI-RAN), which implements a 4G and 5G 
(eNB, gNB, and 4G/5G UE) RAN, and OAI Core Network, which implements the Evolved Packet Core 4G and 5G Core Network.
The performance metric is the $P_\textrm{d}$ assuming $P_\textrm{fa}=10^{-2}$, whereas $K$ depends on the specific
scenario. Finally, the OAI-RAN, emulated by the USRP, follows the release 15 of the 3GPP standard.
\vspace{-0.20cm}
\subsection{Operating Scenarios}
\begin{figure}[h!]
\centering
\includegraphics[width=7.5cm,height=4.75cm]{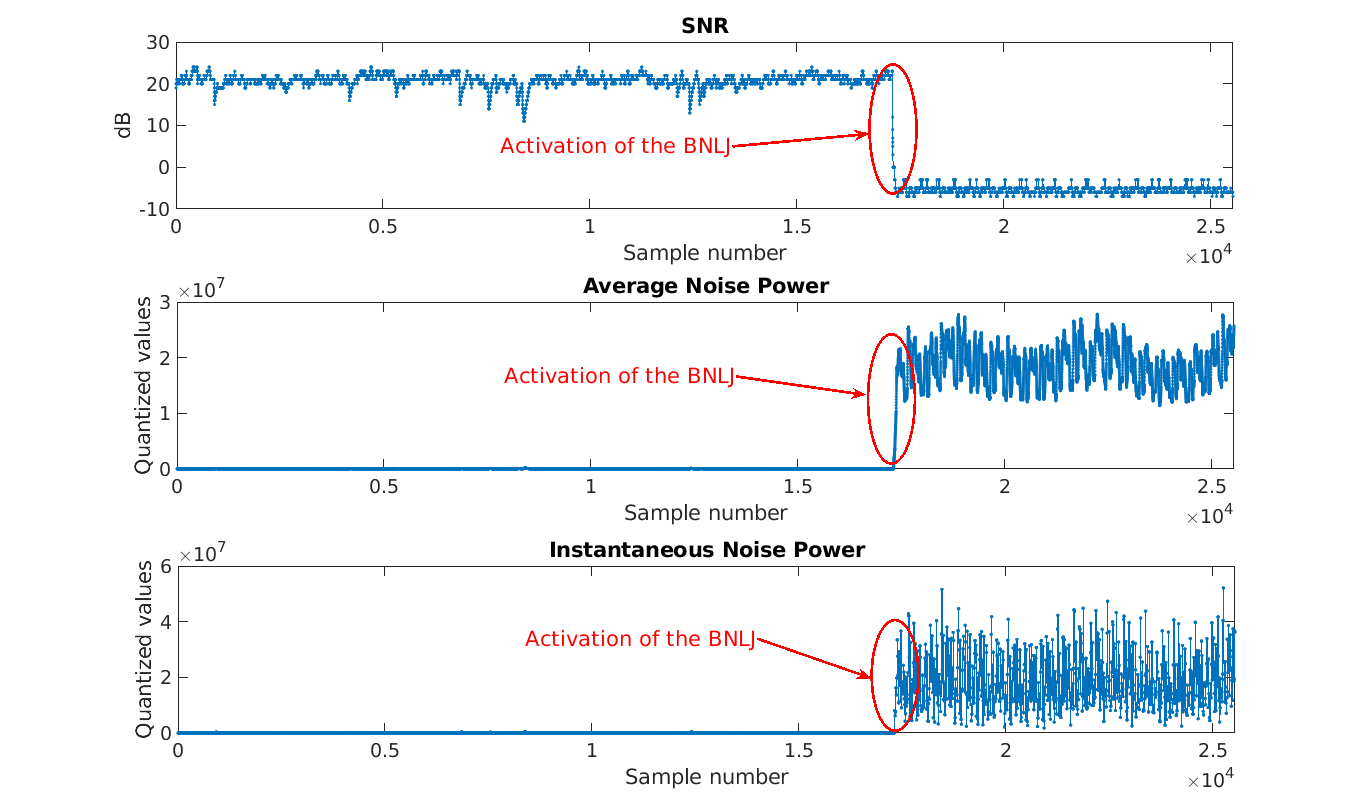}
\caption{SNR, average noise power, and instantaneous noise power versus sample number for the scenario that includes a BNLJ 
with $\mbox{J/S}=5$ dB.}
\label{fig:dataBNLJ_25dB}
\end{figure}

\begin{figure}[h!]
\centering
\includegraphics[width=7cm,height=4cm]{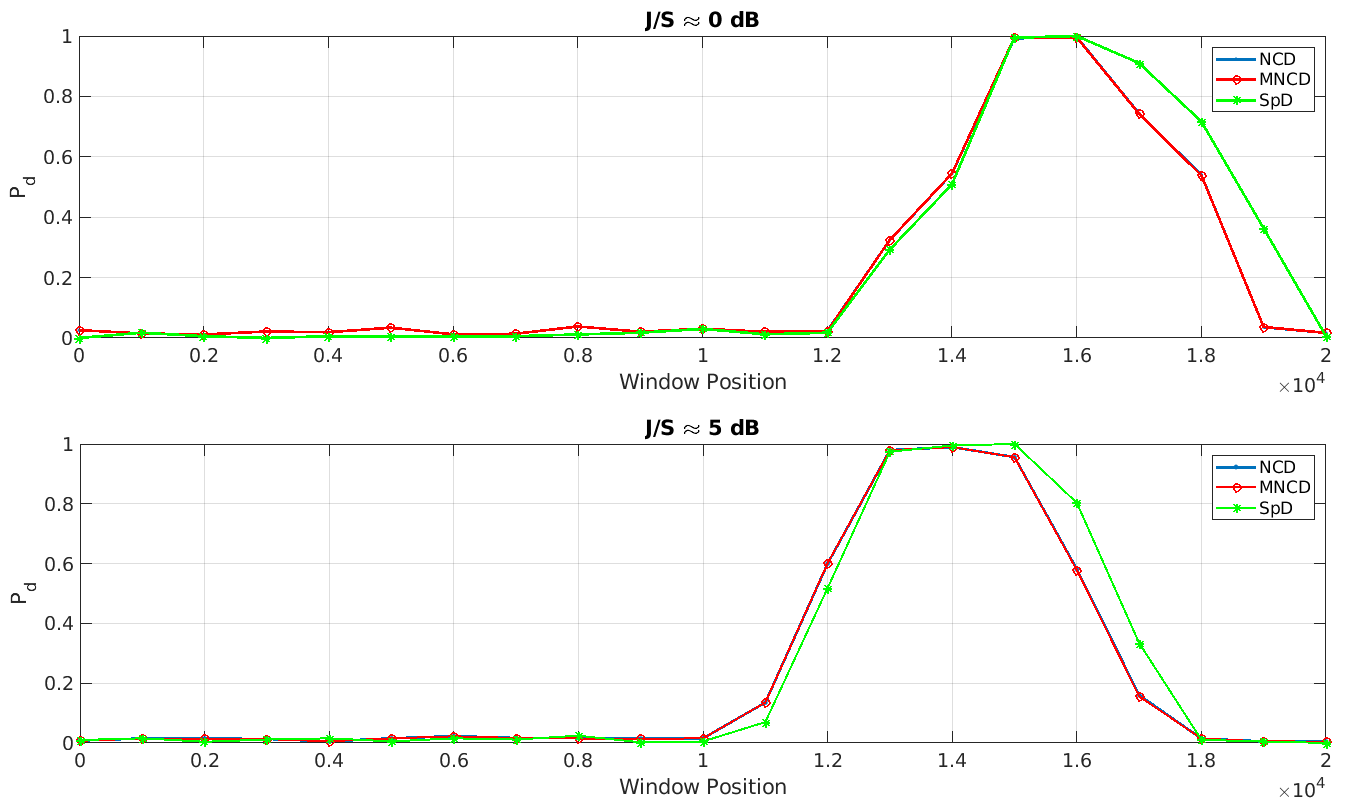}
\caption{$P_\textrm{d}$ versus sliding window position for the scenario that includes a BNLJ with J/S equal to $0$ and $5$ dB.}
\label{fig:Pd_BNLJ_25dB}
\end{figure}

\begin{figure}[h!]
\centering
\includegraphics[width=6cm]{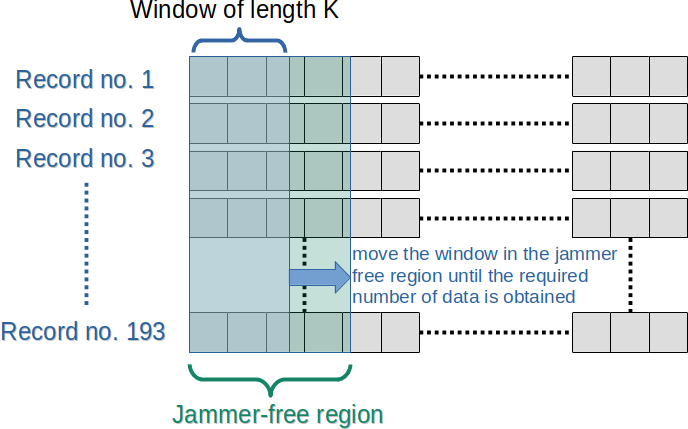}
\caption{Data selection procedure for threshold estimation applied to the first operating scenario.}
\label{fig:thresholdData}
\end{figure}

\paragraph*{Scenario 1} In the first operating scenario, the threat is represented by a 
BNLJ illuminating the the entire communication channel used by UE in 
order to saturate its receiver forcing the disconnection from the genuine base station. 
In Figure \ref{fig:dataBNLJ_25dB}, we show the effects of the implemented BNLJ on the collected measurements.
As expected, the noisy signals injected into the UE receiver lead to an abrupt variation of both the measured noise level and the SNR.
The detection performance of the proposed architectures in the presence of the BNLJ attack 
with $\mbox{J/S}$ equal to $0$ and $5$ dB is shown in Figure \ref{fig:Pd_BNLJ_25dB} 
using $2$-dimensional vectors containing the SNR and the average noise power. 
The overall data matrix contains $173$ ($193$) rows (representative of the number
of records) and $26809$ ($25535$) columns (namely, the number of samples collected for each record) for $\mbox{J/S}=0$ dB
($5$ dB), whereas the length of the temporal
sliding window is $K=5001$ samples. The undersampling of a factor $10$ is also performed before applying the algorithms.
The detection threshold is estimated through Monte Carlo simulation where data are generated 
by selecting $K$ samples from the initial part 
of each record that is jammer-free. Specifically, we process $193/P_\textrm{fa}$ windows 
of length $K$ generated as described in Figure \ref{fig:thresholdData}.
This analysis highlights that NCD, MNCD, and SpD share good detection performance and 
can declare the presence of a BNLJ when the latter begins
its transmission. It is clear that once the BNLJ has modified the measurement values the $P_\textrm{d}$ curves drop close to zero
due to the stationarity of the interference effects.

\begin{figure}[tbp]
\centering
\includegraphics[width=7.5cm,height=4cm]{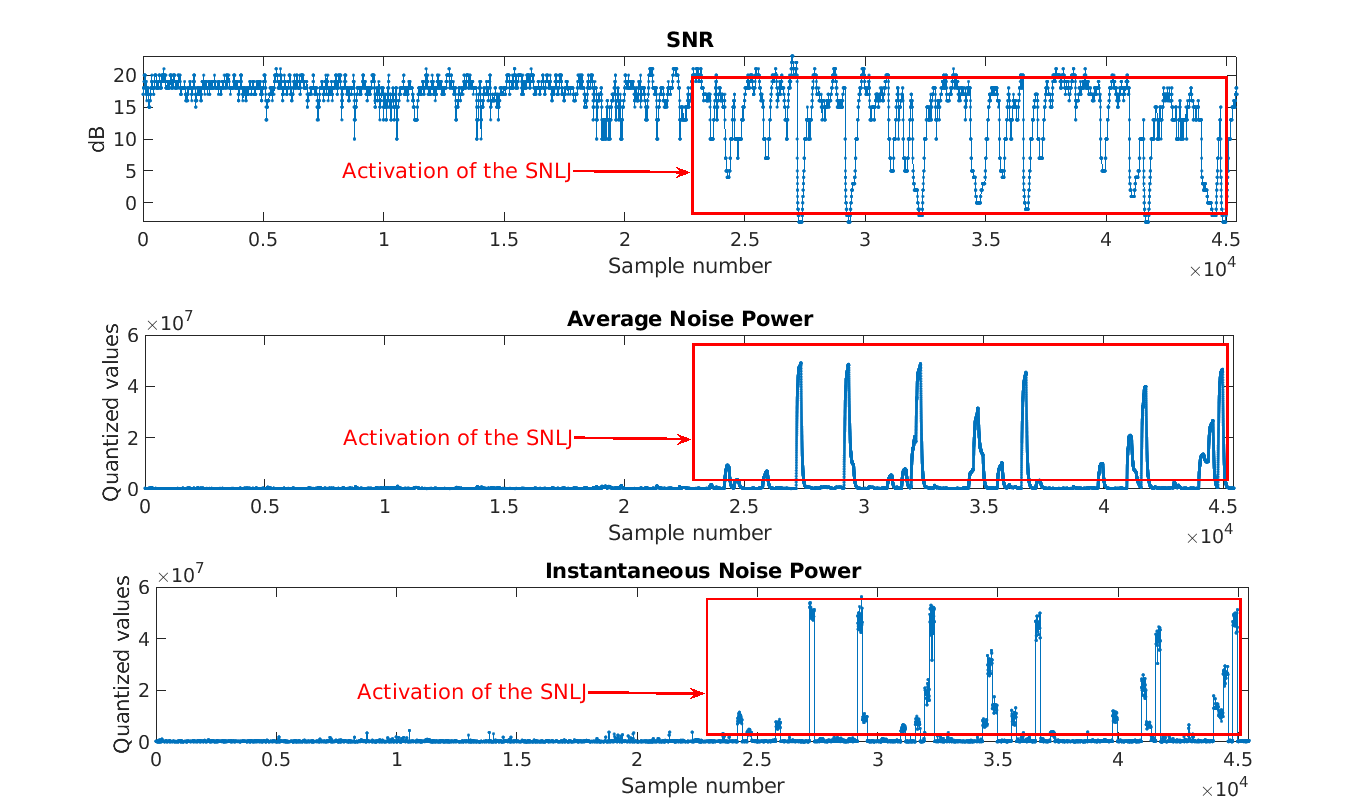}
\caption{SNR, average noise power, and instantaneous noise power versus sample number for the scenario that includes a SNLJ 
with $\mbox{J/S}=5$ dB.}
\label{fig:dataSNLJ_25dB}
\end{figure}

\begin{figure}[tbp]
\centering
\includegraphics[width=6.5cm,height=3cm]{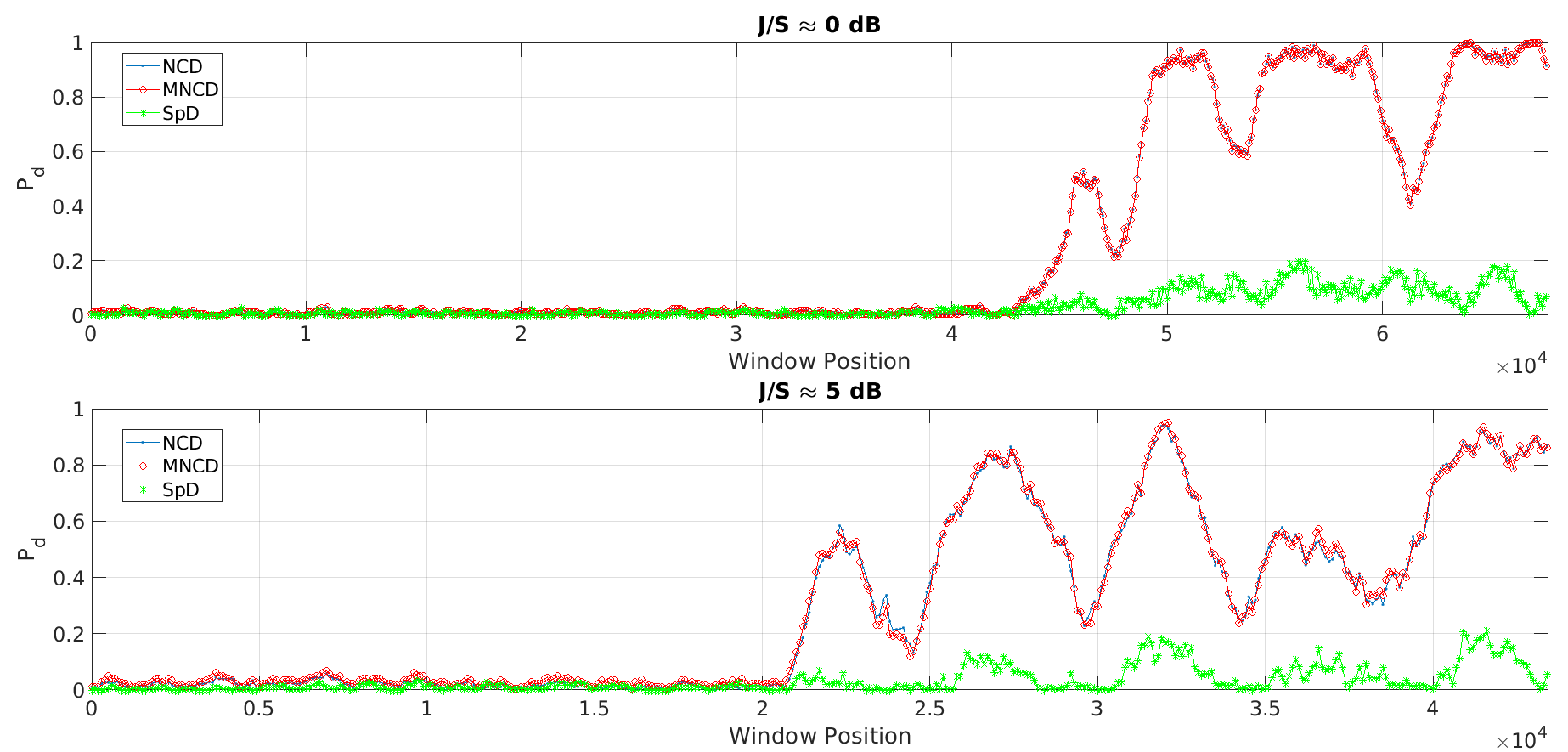}
\caption{$P_\textrm{d}$ versus sliding window position for the scenario that includes a SNLJ with J/S equal to $0$ and $5$ dB.}
\label{fig:Pd_SNLJ_25dB}
\end{figure}

\begin{figure}[tbp]
\centering
\includegraphics[width=6cm,height=2cm]{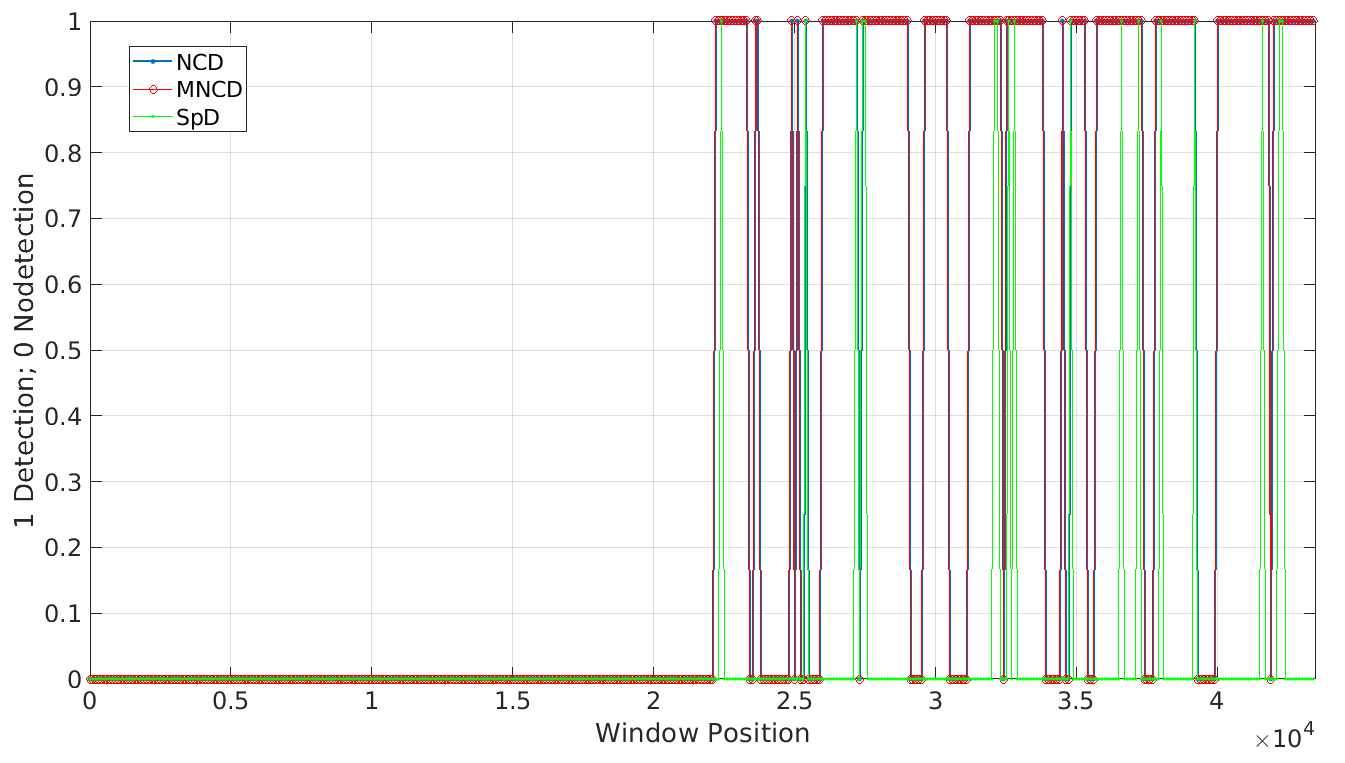}
\caption{Binary function indicating that a detection occurs versus 
sliding window position for the scenario that includes a SNLJ with $\mbox{J/S}=5$ dB.}
\label{fig:Pd_file_SNLJ_25dB}
\end{figure}

\paragraph*{Scenario 2}  The next scenario replaces the BNLJ with a SNLJ maintaining unaltered the J/S, 
whose effects on data can be observed in Figure \ref{fig:dataSNLJ_25dB}. 
The SNLJ does not jam the entire spectrum of the LTE channel, but acts as a frequency hopping jammer. 
Therefore, the jamming signals cover a small window of 60 contiguous sub-carriers (corresponding to 450 KHz of jammed spectrum), 
and these sub-carriers move rapidly within the LTE channel every 0.1 seconds. 
To increase the jamming efficiency, the generated interference signal carries an LTE structure 
with randomly generated LTE pseudo-sub-frames. Finally, it uses the two transmission chains of the Ettus USRP B210 board 
to maximize its effect by sending two independent interference signals, tuned on the same LTE channel but covering 
different spectral windows. The effects of this action consist of a series of spikes in the noise 
power data and notches in the SNR data.
In this scenario, we apply the considered detection architectures using a window of length $K=2001$ with an undersampling factor equal to $10$.
The threshold is set following the same procedure as for the BNLJ with the difference that 
in this case the number of records is $213$ ($178$) and the number of samples is $69641$ ($45442$)
for $\mbox{J/S}=0$ dB ($5$ dB). Figure \ref{fig:Pd_SNLJ_25dB} 
contains the $P_\textrm{d}$ curves obtained by processing average and instantaneous noise power measurements.
The SNR measurements are discarded due to the presence of
fluctuations that are not associated with the SNLJ and lead to a performance degradation.
Note that the NCD and the MNCD share the same performance and can achieve $P_\textrm{d}$ values close to $1$.
On the other hand, the curves related to the SpD are below $P_\textrm{d}=0.5$. Moreover, the figure unveils that an integration rule 
of the {\em contacts} is also required to introduce a hysteresis that mitigates the $P_\textrm{d}$ variations due to the
SNLJ behavior. This need is corroborated by Figure \ref{fig:Pd_file_SNLJ_25dB}, 
where we plot a binary curve that is $1$ if a detection occurs and $0$ otherwise for the case $\mbox{J/S}=5$ dB only.

\begin{figure}[tbp]
\centering
\includegraphics[width=7.5cm,height=5cm]{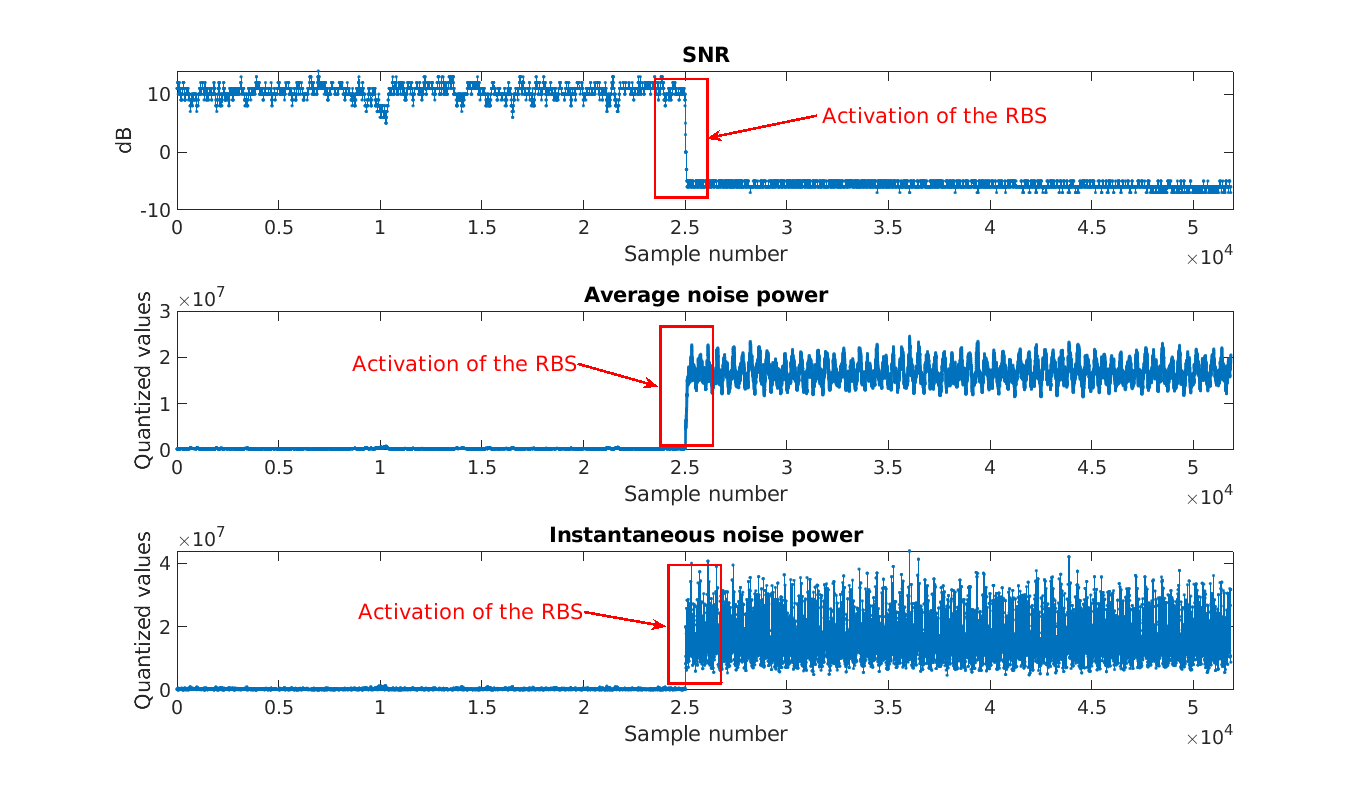}
\caption{SNR, average noise power, and instantaneous noise power versus sample number for the scenario that includes the RBS.}
\label{fig:dataRBS}
\end{figure}

\begin{figure}[tbp]
\centering
\includegraphics[width=6.5cm,height=3.5cm]{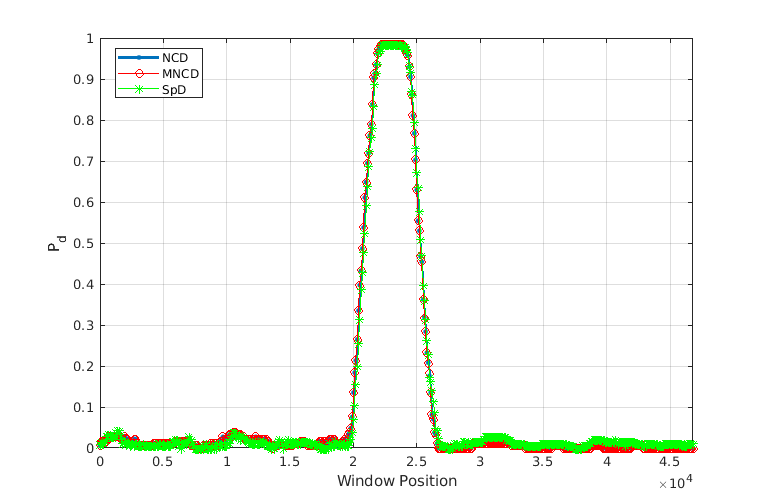}
\caption{$P_\textrm{d}$ versus sliding window position for the scenario that includes the RBS.}
\label{fig:Pd_RBS}
\end{figure}

\paragraph*{Scenario 3} In the last scenario, the RBS comes into play in place of the jammers. RBS signals interfere with those of the genuine base station
as shown in Figure \ref{fig:dataRBS}, where the measurement behavior is similar to that
observed in the presence of the BNLJ. The detection threshold is computed as described for the previous scenarios 
and over $185$ records of $51844$ samples. Vectors $\bz_k$s are formed by using the average and the instantaneous noise power,
hence, $N=2$, whereas $K=5001$. The $P_\textrm{d}$ curves are shown in Figure \ref{fig:Pd_RBS}, where all the considered detection architectures
are capable of achieving $P_\textrm{d}=1$ when the position of the sliding window is such that it contains the time 
instant at which the RBS activates.
In addition, the curves are very close to each other.
 
Summarizing, a window length of about $5000$ samples provides satisfactory performances in the case of BNLJ and RBS. As for the
the SNLJ, the analysis has singled out the NCD and the MNCD with $K=2001$ as the recommended architectures to 
detect the SNLJ. However, 
a further processing layer is required to integrate the number of detections.
Finally, from the computational point of view, the NCD and MNCD share the same complexity that,
in terms of the usual Landau notation, is $\cO(2KN^2+3N^3)$, whereas the SpD is slightly less time demanding
since its computational load is $\cO(2KN^2+2N^3)$.

\vspace{-0.2cm}

\section{Conclusion}
In this letter, we have addressed the problem of detecting the actions of malicious agents aimed at interrupting the communications
between the UE and the legitimate base station to force a connection with the RBS. In this context, we have investigated the effects
of three different threats and conceived detection architectures based upon the GLRT and fed by high-level measurements
provided by the UE. The detection performance have been assessed through real recorded data corresponding
to three different LTE-based scenarios. 
The numerical examples have highlighted that NCD and MNCD can provide satisfactory detection performance in 
all the considered real operating scenarios.

Future research tracks might focus on the design of suitable integration 
techniques to prevent the abrupt $P_\textrm{d}$ variations in the presence of a SNLJ, 
or classification schemes that are capable of recognizing which kind of attack is in progress.

\appendices

\section{Derivation of \eqref{eqn:naiveCD}}
\label{App:derivationNCD}
The decision statistic of the Log-GLRT for problem \eqref{eqn:generalProblem} is
\begin{align}
&\dmax_{K_1}\!\!\dmax_{\bm_i,\bSigma_i \atop i=1,2}\!\!\!
\left\{\sum_{k=1}^{K_1 }
\log f(\bz_k;\bm_1,\bSigma_1)
+\!\!\!\!\!\sum_{k=K_1+1}^{K}\!\!\!\!\log f(\bz_k;\bm_2,\bSigma_2)\right\}  \nonumber
\\
&-\dmax_{K_1}\!\!\dmax_{\bm_i,\bSigma \atop i=1,2}\!\!\!
\left\{
\sum_{k=1}^{K_1 }\log f(\bz_k;\bm_1,\bSigma)
+\!\!\!\!\!\sum_{k=K_1+1}^{K}\!\!\!\!\log f(\bz_k;\bm_2,\bSigma)\right\},
\label{eqn:LLR_01}
\end{align}
where $\log f(\bz;\bm,\bSigma)=-\frac{N}{2}\log(2\pi)-\frac{1}{2}\log\det(\bSigma)
-\frac{1}{2}\tr[\bSigma^{-1}(\bz-\bm)(\bz-\bm)^T]$ is the log-likelihood of a Gaussian 
random vector $\bz$ with mean $\bm$ and covariance matrix $\bSigma$.

The joint maximization with respect to $\bm_i$ and $\bSigma_i$, $i=1,2$, of the first term 
of \eqref{eqn:LLR_01} is tantamount to the following optimization problems
\be
\dmax_{\bm_i,\bSigma_i \atop i=1,2}\Bigg\{ \frac{K_i}{2}\log\det(\bSigma_i^{-1})-\frac{1}{2}\tr\left[\bSigma_i^{-1}
%%%\\
%%%\times
\tilde{\bZ}_i\tilde{\bZ}_i^T \right]\Bigg\},
%(\bZ_i-\bm_i\boe^T)(\bZ_i-\bm_i\boe^T)^T\right]\Bigg\}, \ i=1,2.
\ee
where $\tilde{\bZ}_i=\bZ_i-\bm_i\boe^T$.
It is well known that the maximizers are $\widehat{\bm}_i$ and $\frac{1}{K_i}\bA_i$, $i=1,2$ \cite{muirhead2009aspects}. 
As for the second term of \eqref{eqn:LLR_01}, the maximization problems with respect to $\bm_1$ and $\bm_2$ can be 
separately solved leading to $\widehat{\bm}_i$, $i=1,2$. It follows that the last maximization problem with respect to $\bSigma$ is
\be
\dmax_{\bSigma}\frac{K}{2}\left\{\log\det(\bSigma^{-1})-\tr\left[ \bSigma^{-1}(\bA_1+\bA_2)/K \right]\right\},
\label{eqn:maxH0}
\ee
which can be solved using the inequality $\log\det(\bC)\leq \tr(\bC)-N$ with $\bC$ any positive 
definite $N$-dimensional matrix \cite{lutkepohl1996handbook}.
Therefore, the solution of \eqref{eqn:maxH0} is $\widehat{\bSigma}=(\bA_1+\bA_2)/K$ and the proof is complete.

\section{Derivation of \eqref{eqn:modifiedNCD}}
\label{App:derivationMNCD}
The decision statistic of the Log-GLRT for problem \eqref{eqn:sameMeanVarianceH0} is
\begin{align}
\dmax_{K_1}\!\!\dmax_{\bm_i,\bSigma_i \atop i=1,2}\!\!\!
&\left\{\sum_{k=1}^{K_1 }
\log f(\bz_k;\bm_1,\bSigma_1)
+\!\!\!\!\!\sum_{k=K_1+1}^{K}\!\!\!\!\log f(\bz_k;\bm_2,\bSigma_2)\right\}  \nonumber
\\
&-\dmax_{\bSigma, \bm}
\sum_{k=1}^{K}\log f(\bz_k;\bm,\bSigma).
\label{eqn:LLR_02}
\end{align}
Therefore, the maximization problem in the first term can be solved by resorting 
to the results of Appendix \ref{App:derivationNCD}, whereas the second
term leads to the well-known maximum likelihood estimates of $\bm$ and $\bSigma$ under $H_0$ given 
by $\widehat{\bm}$ and $\frac{1}{K}\bA_0$, respectively.

\bibliographystyle{IEEEtran}
\bibliography{group_bib}

\end{document}